\documentclass[twocolumn,aps,prd,amsmath,amssymb,preprintnumbers,longbibliography]{revtex4-1}
\usepackage{amsmath} 
\usepackage{amsfonts} 
\usepackage{amssymb}
\usepackage{tensor}
\usepackage{bbm}
\usepackage{graphics}
\usepackage{graphicx}
\usepackage{titlesec}
\usepackage{mathtools}
\usepackage{environ}
\usepackage{dsfont}

\usepackage[colorlinks=true]{hyperref}
\hypersetup{urlcolor=black,linkcolor=black,citecolor=black}
\usepackage[toc,page]{appendix}

\usepackage{makecell,tabularx}
\setcellgapes{3pt}

\makeatletter
\renewcommand*\env@matrix[1][c]{\hskip -\arraycolsep
  \let\@ifnextchar\new@ifnextchar
  \array{*\c@MaxMatrixCols #1}}
\makeatother

\newcommand{\be}{\begin{equation}}
\newcommand{\ee}{\end{equation}}
\newcommand{\ba}{\begin{eqnarray}}
\newcommand{\ea}{\end{eqnarray}}

\newcommand{\diag}{\text{ diag }}

\titleformat{\subsection}[block]{\normalfont\bfseries}{\thesubsection.}{1ex}{}
\titlespacing{\subsection}{0pt}{10pt}{1pt}[0pt]
\titleformat*{\section}{\large\bfseries}
\renewcommand{\thesubsection}{\arabic{subsection}}

\usepackage{natbib}
\usepackage{braket}

\usepackage{todonotes}

\definecolor{refkey}{rgb}{0,0,1}
\definecolor{labelkey}{rgb}{0,1,0}

\renewcommand{\thesection}{\arabic{section}}
\renewcommand{\thesubsection}{\thesection.\arabic{subsection}}

\titlespacing{\paragraph}{0pt}{15pt}{5pt}[0pt]
\titleformat{\paragraph}[block]{\normalsize\bfseries}{\theparagraph}{1ex}{}

\makeatletter
\renewcommand{\p@subsection}{}
\renewcommand{\p@subsubsection}{}
\makeatother

\newcommand{\rhs}{r.\,h.\,s.}

\newcommand{\tchi}{{\tilde{\chi}}}
\newcommand{\tphi}{{\tilde{\varphi}}}
\newcommand{\trho}{\tilde{\rho}}
\newcommand{\tsigma}{\tilde{\sigma}}
\newcommand{\tpsi}{\tilde{\psi}}
\newcommand{\tJ}{\tilde{J}}
\newcommand{\p}{\partial}

\DeclareMathOperator{\tr}{tr}

\newcommand{\Gammatwo}{\Gamma^{(2)}}
\renewcommand{\phi}{\varphi}
\newcommand{\inv}[1]{#1^{-1}}
\newcommand{\comm}[1]{\left[#1 \right]}
\newcommand{\anticomm}[1]{\left\{#1 \right\}}
\newcommand{\hc}{\mathrm{h.\,c.}}
\renewcommand{\epsilon}{\varepsilon}
\DeclareMathOperator{\Lapl}{\mathcal{D}}

\newcounter{aufz}
\newcommand{\aufz}{\addtocounter{aufz}{1}(\arabic{aufz})~}

\begin{document}

\title[ ]{Fundamental Scale Invariance}

\author{C. Wetterich}
\affiliation{Institut  f\"ur Theoretische Physik\\
Universit\"at Heidelberg\\
Philosophenweg 16, D-69120 Heidelberg}

\begin{abstract}
We propose fundamental scale invariance as a new theoretical principle beyond renormalizability.
Quantum field theories with fundamental scale invariance admit a scale-free formulation of the functional integral and effective action in terms of scale invariant fields. They correspond to exact scaling solutions of functional renormalization flow equations. Such theories are highly predictive since all relevant parameters for deviations from the exact scaling solution vanish. 
Realistic particle physics and quantum gravity are compatible with this setting.
The non-linear restrictions for scaling solutions can explain properties 
as an asymptotically vanishing cosmological constant or dynamical dark energy
that would seem to need fine tuning of parameters from a perturbative viewpoint.
As an example we discuss a pregeometry based on a diffeomorphism invariant Yang-Mills theory. It is a candidate for an ultraviolet completion of quantum gravity with a well behaved graviton propagator at short distances.
\end{abstract}

\maketitle

The understanding of scale symmetry and its possible breaking is a central issue in quantum field theories. It has been discussed in the context of an understanding of the gauge hierarchy in particle physics\,\cite{CWFT,BAR,HEM,NIME,FKV,AOI,WEYA}, the cosmological constant\,\cite{RAB,CWQ} or inflation\,\cite{SUG,INF1,CWCI,CWVG,CWQIM,INF2,RUW}.
Quantum scale symmetry\,\cite{CWQS} is a powerful symmetry that can render a model highly predictive. 
It requires that the effective action remains invariant under suitable multiplicative rescalings of the renormalized fields.
In this note we are motivated by three observations for which we develop a unified view. 

\aufz A quantum scale symmetric standard model has been proposed in refs.\,\cite{CWQ,SH}. For such a model the quantum effective action does not contain any intrinsic parameter with dimension of length or mass. This distinguishes quantum scale symmetry from models with classical symmetry\,\cite{CSS1,CSS2,CSS3,CSS4,CSS5,CSS6}. For a scale invariant classical action (classical scale symmetry) the running of couplings due to quantum fluctuations typically introduces explicit mass scales and violates scale symmetry. The quantum scale invariant standard model introduces an additional scalar singlet $\chi$. The running of couplings occurs now as functions of $q^2/\chi^2$ or $h^\dagger h/\chi^2$, with $q^2$ the squared momentum and $h$ the field for the Higgs doublet. Dimensionless ratios involving intrinsic mass scales are replaced by ratios of field values or ratios between momenta and fields. No intrinsic mass or length is present. This points to a fundamental theory without scales, where the running of effective couplings arises through their dependence on fields.

\aufz Scale symmetry is spontaneously broken whenever a scalar field takes a non-zero value. This is the case for the Higgs doublet or for the scalar field $\chi$ that replaces the Planck mass. In case of spontaneously broken exact scale symmetry one expects the presence of an exactly massless Goldstone boson. The proposal of dynamical dark energy\,\cite{CWQ} is based on a small dilatation anomaly. A scale invariant coupling to gravity replaces the Planck mass $M$ by a scalar field $\chi$, such that the curvature scalar $R$ appears in the effective action in the form
\begin{equation}
\Gamma = \int_x \sqrt{g} \left\{ -\frac{1}{2} \chi^2 R + ck^4 \right\}.
\label{eq:I1}
\end{equation}
Here $g$ is the determinant of the metric $g_{\mu\nu}$.
For $c\neq 0$ scale symmetry is explicitly broken by the scale $k$ which has dimension of mass. The effective cosmological constant corresponds to the dimensionless ratio of scalar potential over the fourth power of the Planck mass, which is given for eq.\,\eqref{eq:I1} by
\begin{equation}
\lambda = \frac{ck^4}{\chi^4}.
\label{eq:I2}
\end{equation}

For cosmological solutions $\chi$ is found to increase without bounds such that the cosmological constant vanishes asymptotically in the infinite future. At present, the Universe is old, but not infinitely old. The scalar field $\chi$ still has a finite value at the present time $t_0$, for which we may use the units $\chi(t_0)=M$, with $M$ the (reduced) Planck mass. For a large ratio $\chi(t_0)/k$ one expects a small amount of dark energy, which is dynamical since $\chi$ increases with time. This early prediction of dynamical dark energy\,\cite{CWQ} seems to point towards a small explicit breaking of quantum scale symmetry by the scale $k$, which is of the order $10^{-3}\,\mathrm{eV}$ for units with $\chi(t_0)=M$. The Goldstone boson becomes a pseudo-Goldstone boson -- the cosmon. This is the almost massless field of dynamical dark energy. (In scale invariant unimodular gravity the term $ck^4$ arises as an integration constant rather than as an intrinsic parameter\,\cite{SUG}. Since the predictions for observations are identical to the explicit breaking in the effective action \eqref{eq:I1}, one finds again a pseudo-Goldstone boson. In view of the absence of an exact Goldstone boson the interpretation as a spontaneously broken exact global scale symmetry is not clear to us.)

\aufz Scaling solutions for flow equations\,\cite{CWFE,CWMR,MR} have been investigated for dilaton quantum gravity\,\cite{DG1,DG2}. This generalizes the effective action \eqref{eq:I1},
\begin{align}
\begin{split}
\Gamma = \int_x \sqrt{g} \left\{ -\frac{1}{2} F(\chi) R + U(\chi) + \frac{1}{2} K(\chi) \p^\mu \chi \p_\mu \chi \right\},
\end{split}
\label{eq:I3}
\end{align}
where the three functions $F$, $U$ and $K$ flow with 
a renormalization scale $k$. For scaling solutions of functional flow equations the functions $F/k^2$, $U/k^4$ and $K$ only depend on the ratio $\chi^2/k^2$, without the presence of any other mass scale.
The candidate scaling solutions found show indeed for large $\chi$ the behavior of the effective action \eqref{eq:I1}. A general investigation of scaling solutions for effective potentials\,\cite{ESPA} finds scaling potentials that approach constants for large field values, with a limit \eqref{eq:I1}. 

The scaling
functions $U/k^4$, $F/k^2$, $K$
depend on dimensionless field ratios as $\trho = \chi^2/(2k^2)$. The effective action \eqref{eq:I3} therefore involves a scale $k$. It has been observed\,\cite{DG2,ESPA} that the scale $k$ disappears when the model is transformed by a Weyl scaling of the metric to the Einstein frame. This suggests that $k$ may actually not play the role of an intrinsic parameter with dimension of mass. Combined with a scale invariant standard model, for which all mass scales are proportional to $\chi$, the quantum field equations derived from the effective action \eqref{eq:I3} are the ones of variable gravity\,\cite{CWVG}. Rather realistic cosmologies are obtained in this context\,\cite{CWQIM,RUW}. Thus the scaling solutions of flow equations may result in an acceptable cosmology and particle physics, without any need that the flow deviates from the scaling solutions due to some relevant parameters.

In the present note we develop a coherent picture of these three facets
of quantum scale symmetry. 
They seem at first sight a bit contradictory. The quantum scale invariant standard model points towards \textit{exact} quantum scale symmetry, while dynamical dark energy seems to suggest 
only \textit{approximate} quantum scale symmetry, with
a breaking by an intrinsic scale $k$. One point of view observes that $k\approx 10^{-3}\,\mathrm{eV}$ is much smaller than the scales relevant for the standard model, such that the tiny scale anomaly is actually negligible except for the scales of present cosmology. For all other scales one may think that the small scale anomaly $\sim k^4$ plays no role, rendering models with $k\neq 0$ indistinguishable from models with exact scale symmetry and $k=0$. This argument 
is valid for the quantum scale invariant standard model. It
does not hold, however, for very early cosmology, as the inflationary epoch. For cosmon inflation\,\cite{CWCI,CWVG,CWQIM,RUW} the cosmon field $\chi$ is smaller than $k$ for the very early epochs of inflation. It is the scale $k$ that triggers the end of inflation once $\chi$ increases sufficiently beyond $k$. The same scale is also responsible for the small deviations from scale invariance of the primordial fluctuation spectrum. Thus again the scale $k$ plays a useful role. 

Combined with the possibility to eliminate $k$ by a transition to the Einstein frame, the successful cosmology suggests the presence of a ``renormalization scale'' $k$ which appears in scaling solutions of flow equations.
In this note we will develop a deeper view for which the scale $k$ is actually present, but does not correspond to an intrinsic parameter violating scale invariance explicitly.

\section*{Scale invariance}

We propose that a fundamental theory has no scale. More precisely, a fundamental quantum field theory does not involve any intrinsic parameter with dimension mass or length. 
This is the meaning of ``fundamental scale invariance''.
The fields for the most basic constituents are dimensionless. Fields depend on spacetime coordinates, and one may decide to introduce a unit of length for distances between spacetime points. Correspondingly, derivatives of fields with respect to the spacetime coordinates or momenta carry dimension of inverse length or mass. (We use units $\hbar=c=1$.) A metric field may arise as a composite or collective field constructed from derivatives of fundamental fields $\tpsi$
\begin{equation}
\tilde{g}_{\mu\nu} \sim f(\tpsi) \p_\mu \tpsi \p_\nu \tpsi,
\label{eq:I4}
\end{equation}
where we do not spell out other possible indices or the form of $f$. This metric has therefore dimension mass squared. 

Geometry is usually constructed with a dimensionless metric. For this purpose one introduces a scale $k$ with dimension mass,
\begin{equation}
g_{\mu\nu} = k^{-2} \tilde{g}_{\mu\nu}.
\label{eq:I5}
\end{equation}
Scalars are either fundamental fields, or composites of fundamental fields. Involving no derivatives, scalar fields $\tchi$ are dimensionless on a fundamental level. We may decide to use a different ``canonical'' normalization
\begin{equation}
\chi=k\tchi,
\label{eq:I6}
\end{equation}
such that $\chi$ carries dimension of mass and a diffeomorphism invariant kinetic term can be canonical. 

Obviously, the scale $k$ has no physical meaning and is not an intrinsic parameter. It is introduced only for convenience. The model could be formulated in terms of fields as $\tpsi$, $\tchi$, $\tilde{g}_{\mu\nu}$, for which the scale $k$ never appears. The quantum effective action, formulated in terms of fields as $\tpsi$, $\tchi$, $\tilde{g}_{\mu\nu}$, is trivially independent of $k$. We call the fields $\tpsi$, $\tchi$, $\tilde{g}_{\mu\nu}$ ``scale invariant fields'', since they are associated to a formulation for which no scale appears. The rescaled fields as $g_{\mu\nu}$ or $\chi$ may be called ``canonical fields''.
We stress the difference between scale and dimension. Dimensions of space distances, derivatives, momentum and fields appear once one decides to associate a dimension of length to space distances. Scales or ``scaling dimensions'' for fields appear once one uses canonical fields as in eqs.\,\eqref{eq:I5}, \eqref{eq:I6}. 
The example of the metric $\tilde{g}_{\mu\nu}$ shows that scale invariant composite fields can carry dimension.

Expressed in terms of the canonical fields the effective action will generically depend on $k$
\begin{equation}
k\p_k \Gamma_k[\phi] = \zeta_k[\phi],
\label{eq:I7}
\end{equation}
with $\phi$ standing collectively for canonical fields as $g_{\mu\nu}$ and $\chi$. The flow generator $\zeta_k[\phi]$ does not vanish, and the flow equation \eqref{eq:I7} describes the dependence of the effective action on the scale $k$. On the other hand, we know that for fixed scale invariant fields $\tphi$ the effective action does not involve $k$,
\begin{equation}
k \p_k \Gamma_k [\tphi] = 0.
\label{eq:I8}
\end{equation}
The general solutions of the differential equation \eqref{eq:I7} therefore include a particular scaling solution for which eq.\,\eqref{eq:I8} holds once the canonical fields are expressed in terms of the scale invariant fields. It is precisely this scaling solution that defines the theory with fundamental scale invariance. It expresses the fact that the dependence on $k$ is introduced into the model only by a redefinition of fields \eqref{eq:I5}, \eqref{eq:I6}. So far these statements seem almost trivial, related to field rescalings rather than running couplings. We will see that they continue to hold for situations where the flow generator $\zeta_k$ describes the physical effects of running couplings.

\section*{Quantum field theories without scale}

Let us consider some well-defined (regularized) quantum field theory involving dimensionless fields $\tsigma_i(x)$. Here $x$ may be the sites of a discrete lattice or the space-time points of a continuous manifold. Examples are lattice gauge theories with $\tsigma$ the link variables, or lattice spinor gravity\,\cite{CWLSG} with Grassmann variables $\tsigma$ describing ``fundamental fermions''. A given quantum field theory is specified by a functional integral over the fields $\tsigma$, with an action $S[\tsigma]$ being a functional of these fields.

The quantum effective action $\Gamma [\tphi]$ is defined by a functional differential equation (``background field identity'')
\begin{equation}
\exp(-\Gamma[\tphi]) = \int D\tchi \exp \left\{ -S[\tphi+\tchi] + \int_x \frac{\p\Gamma}{\p\tphi} \tchi \right\}.
\label{eq:NS1}
\end{equation}
This effective action is a functional of the multicomponent macroscopic fields $\tphi_i(x)$, treated here as a vector $\tphi$. It involves the (euclidean) action $S[\tphi+\tchi]$, which is a functional of the microscopic fields $\tsigma=\tphi+\tchi$. The functional integration over $\tilde{\sigma}$ is shifted to an integral over the fluctuation fields $\tchi$. The first functional derivatives of $\Gamma$ are called sources
\begin{equation}
\tJ_i(x) = \frac{\p \Gamma}{\p \tphi_i(x)},\quad \tilde{J} = \frac{\p\Gamma}{\p\tphi},
\label{eq:NS2}
\end{equation}
and $\int_x \tJ\tchi$ is the scalar product of the source vector and the fluctuation vector. For fermions $\tphi$ and $\tchi$ are Grassmann variables. For a continuum formulation of local gauge theories one adds a gauge fixing term and the associated Faddeev-Popov determinant or ghost term. A gauge invariant effective action can be obtained by a ``physical gauge fixing''\,\cite{CWGIF}.
The effective action is the generating functional for the one-particle irreducible Green's functions. All information relevant for observations can be extracted from its functional derivatives. The first derivative yields the field equations in the presence of the sources \eqref{eq:NS2}, and the second derivative $\Gammatwo$ defines the inverse propagator. Evaluating the propagator on a solution of the field equations yields the fluctuation spectrum. For example, the primordial fluctuation spectrum in inflationary cosmology can be directly extracted from $\Gammatwo$\,\cite{CWCFS}.

Let us focus on a discretized theory, formulated on a lattice. We denote a typical distance between lattice points by $a$, and consider physical phenomena involving distances $l$ of many lattice points, $l/a \gg 1$. The units for $a$ or $l$ do not matter, what only counts is the ratio $l/a$. For example, we could choose $a=1$ or define a length unit by some multiple of $a$. We are interested in the continuum limit $l/a\to\infty$. 
For this purpose we keep $a$ fixed, and consider fixed parameters of the lattice model. 
For a theory with fundamental scale invariance we
require that the expectation values of observables of interest, which can be constructed from suitable correlation functions, have a well defined limit for $l\to\infty$. (For $l\to\infty$ expectation values can either diverge or reach finite values, including zero -- we disregard here the logical possibility of limit cycles.)

In other words, theories with fundamental scale invariance admit meaningful observables that reach for $l\to\infty$ values that remain finite and do not all tend to zero. The effective action $\Gamma[\tphi]$ remains well defined in the continuum limit.
Fundamental scale invariance
is a highly non-trivial property. For the example of lattice-QCD for strong interactions this requirement is not met. 
For the continuum limit of QCD one has to adjust parameters of the lattice theory in order to keep fixed observables at $l$ for $l/a$ increasing. The parameters of the microscopic lattice theory depend on $l/a$. This is not compatible with fundamental scale invariance for which parameters are fixed.
On the other hand, any finite theory obeys this condition. We can keep $l$ fixed and move $a\to 0$. For a finite theory all quantities relevant for observation remain finite in this limit.

For a given effective action one may ask what singles out fundamental scale invariance among other renormalizable theories.
For general renormalizable theories
the effective action remains well defined in the continuum limit if one employs renormalized fields $\phi_{\mathrm{R},i}(x)$. They are related to possible scale invariant fields $\tphi_i(x)$ by use of a renormalization scale $k$,
\begin{equation}
\phi_{\mathrm{R},i}(x) = k^{d_i} f_i(k) \tphi_i(x),
\label{eq:NS3}
\end{equation}
with $d_i$ defining the scaling dimensions of $\phi_{\mathrm{R},i}(x)$ and $f_i(k)$ some possible dimensionless functions of $k$ which gives rise to so called anomalous dimensions. The renormalization scale $k$ has dimension of $\inv{a}$, typically mass or inverse length. Any non-constant $f_i(k)$ needs to involve some other mass scale, often given by $f_i(ka)$. For renormalizable theories $\Gamma[\phi_\mathrm{R}]$ remains finite in the continuum limit taken at fixed $\phi_\mathrm{R}$. Possible divergences at fixed $\tphi$ are then connected to the relation \eqref{eq:NS3} between $\phi_\mathrm{R}$ and $\tphi$. 
The factors $k^{d_i}$ may cancel between different fields. Usually, such cancellations do not happen for anomalous dimensions for which divergences
can appear for $ka\to 0$. 
Also dimensionless renormalized couplings may depend on $ka$. For general renormalizable theories, 
the existence of a continuum limit requires 
that for $ka\to 0$ the dimensionless couplings reach ultraviolet fixed point values. The flow away from the fixed point typically involves a dependence on $ka$, and therefore on $k$ if $a$ is kept fixed.

Theories with fundamental scale invariance are renormalizable theories with the additional property that 
any dependence of the effective action on a
renormalization scale $k$ 
can be absorbed into a definition of scaling fields.
This is the case for finite theories, but the class of theories with well defined effective action $\Gamma[\tphi]$ may be larger. We can define theories with fundamental scale symmetry by the property \eqref{eq:I8}.
In other words, if one can find a choice of fields $\tilde{\varphi}$ for which all mass scales are eliminated in the effective action, this indicates fundamental scale symmetry. For a theory defined in terms of dimensionless fundamental fields the
absence of a dependence on $k$ is trivial if no renormalization scale is introduced. The non-trivial part consists in the statement that $\Gamma[\tphi]$ is well defined in the continuum limit.
From the point of view of the macroscopic theory as encoded in $\Gamma$ a possible choice of fields realizing eq.\,\eqref{eq:I8} constitutes a bridge to a microscopic fundamental theory without scale. It indicates that a continuum limit with fixed lattice couplings is possible.

Eq.\,\eqref{eq:I8} is easily translated to the $k$-dependence of the effective action at fixed renormalized fields $\phi_{\mathrm{R},i}(x)$,
\begin{equation}
\p_k \Gamma[\phi_\mathrm{R}] + \int_x \sum_i \frac{\p\Gamma}{\p\phi_{\mathrm{R},i}(x)} \left. \p_k \phi_{\mathrm{R},i}(x) \right|_\tphi = 0,
\label{eq:NS6}
\end{equation}
or
\begin{equation}
k\p_k \Gamma[\phi_\mathrm{R}] = -\int_x \sum_i \left( d_i + \frac{\p \ln f_i}{\p\ln k} \right) \phi_{\mathrm{R},i}(x) \frac{\p\Gamma}{\p\phi_{\mathrm{R},i}(x)} .
\label{eq:NS7}
\end{equation}
The $k$-dependence or the ``flow'' of the effective action at fixed renormalized fields is non-trivial for theories with fundamental scale invariance. 
This also holds for fixed canonical fields $\varphi$ for which we set $f_i = \mathrm{const.}$ in eq.\,\eqref{eq:NS7}.

In order to judge if eq.\,\eqref{eq:NS7} is obeyed we need an independent identity for $\partial_k \Gamma[\varphi]$. This is provided by the exact flow equation to which we will turn below. This flow equation yields an expression for the flow generator $\zeta_k$ in eq.\,\eqref{eq:I7} which has a one-loop form.
Solutions of this flow equation obeying eq.\,\eqref{eq:NS7} are ``scaling solutions''. 
If a scaling solution can be found the condition for a theory with fundamental scale invariance is met.

\section*{Scale invariant fields for quantum gravity}

The introduction of renormalized or canonical fields is not mandatory for theories with fundamental scale invariance. It is, however, often very convenient. An example is a metric that arises as a composite field
\begin{equation}
\tilde{g}_{\mu\nu} (x) = \p_\mu \tilde{H}_a(x) \p_\nu \tilde{H}_b(x) G^{ab},
\label{eq:NS8}
\end{equation}
with $\tilde{H}_a(x)$ some combinations of dimensionless fundamental fields $\tilde{\psi}_i(x)$, and summations over double indices implied. Due to the derivatives, this metric has dimension mass squared. For a description of geometry one would like to introduce a dimensionless metric, and may do so by using $g_{\mu\nu}(x) = \tilde{g}_{\mu\nu}(x)/k^2$.
For quantum gravity the renormalized field is the canonical field. Since the canonical dimensions for geometric quantities are one of the main motivations for the use of canonical fields, 
we will next describe the scale invariant fields and the notion of fundamental scale invariance in some more detail for quantum gravity, starting from a formulation with canonical fields.

Consider an effective action for the metric $g_{\mu\nu}$ and a scalar field $\chi$ of the form
\eqref{eq:I3} with
effective potential
\begin{equation}
U(\chi) = \frac{\mu^2}{2} \chi^2 + \frac{1}{8} \delta(\chi) \chi^4.
\label{eq:SF2}
\end{equation}
The scale invariant metric and scalar field are given by
\begin{equation}
\tilde{g}_{\mu\nu} = k^2 g_{\mu\nu},\quad \tilde{\chi} = \frac{\chi}{k},\quad \trho = \frac{1}{2} \tilde{\chi}^2.
\label{eq:SF3}
\end{equation}
In terms of these fields the effective action \eqref{eq:I3} reads
\begin{equation}
\Gamma = \int_x \sqrt{\tilde{g}} \left\{ -w \tilde{R} + \frac{1}{2} K \p_\mu \tilde{\chi} \p_\nu \tilde{\chi} \tilde{g}^{\mu\nu} + u \right\},
\label{eq:SF4}
\end{equation}
with
\begin{equation}
w = \frac{F}{2k^2},\quad u = \frac{U}{k^4},
\label{eq:SF5}
\end{equation}
and $\tilde{R}$ the curvature scalar for the metric $\tilde{g}_{\mu\nu}$.

This effective action is independent of $k$ if $w$, $u$ and $K$ only depend on $\tilde{\chi}$ or the invariant $\trho$. This is precisely the case for the scaling solution of flow equations. In general, the requirement of independence of $k$ constitutes a strong restriction. If $F$ contains an intrinsic mass scale as the Planck mass $M$, for example $F = M^2 + 2w_0 k^2 + \xi \chi^2/2$, the function $w$ involves the ratio $M^2/k^2$ and therefore depends on $k$
\begin{equation}
w = \frac{M^2}{2k^2} + w_0 + \frac{\xi}{2} \trho.
\label{eq:SF6}
\end{equation}
Only for $M^2 = 0$ the effective action for the scale invariant fields does not involve $k$. Similarly, for
\begin{equation}
u = \frac{\mu^2}{k^2} \trho + \frac{\delta}{2} \trho^2,
\label{eq:SF7}
\end{equation}
the independence of $k$ requires $\mu^2 = 0$ and $\delta$ to depend only on $\trho$. From the point of view of flow equations the parameters as $M^2$ or $\mu^2$ denote deviations from the scaling solution due to relevant parameters
near an ultraviolet fixed point.
The condition that the effective action is independent of $k$, once it is expressed in terms of the scale invariant fields, requires that it corresponds precisely to the scaling solution of flow equations. 
Fundamental scale invariance predicts the vanishing of all relevant parameters as $M^2$ or $\mu^2$.

Scaling solutions are subject to non-linear differential equations, see below. This severely restricts the possible form of scaling functions as $u(\tilde{\rho})$ or $w(\tilde{\rho})$. Within the setting \eqref{eq:I3} the scaling solutions found so far\,\cite{DG1,DG2,ESPA} imply for $w(\tilde{\rho}\to\infty)$ the behavior \eqref{eq:SF6} with $\xi\neq 0$, $M^2=0$, while $u(\tilde{\rho})$ tends to a constant, $u(\tilde{\rho}\to\infty)=u_\infty$. For these scaling solutions the observable cosmological constant, given by the dimensionless ratio
\begin{equation}
\lambda = \frac{U}{F^2} = \frac{u}{4w^2} \to \frac{u_\infty}{\xi^2\tilde{\rho}^2} \to \frac{4u_\infty k^4}{\xi^2 \chi^4},
\label{eq:25A}
\end{equation}
tends to zero in the infinite future for cosmological runaway solutions with $\chi(t\to\infty)\to\infty$. This observation has predicted the presence of dynamical dark energy\,\cite{CWQ}. After Weyl scaling and a standard renormalization of scalar fields the scale $k$ does no longer appear in the effective action.

The scale invariant fields are, in general, not dimensionless. For $\chi$ and $k$ with dimension of mass, and $g_{\mu\nu}$ dimensionless, one finds that $\tilde{\chi}$ is dimensionless, while $\tilde{g}_{\mu\nu}$ carries the dimension of mass squared. If we combine the scale transformations (global dilatation transformations), 
\begin{equation}
\chi \to \alpha\chi,\quad g_{\mu\nu} \to \alpha^{-2} g_{\mu\nu},
\label{eq:SF8}
\end{equation}
with a rescaling of $k$
\begin{equation}
k \to \alpha k,
\label{eq:SF9}
\end{equation}
the scale invariant fields remain indeed unchanged. We emphasize that the notion of scale invariant fields refers to the combined scaling \eqref{eq:SF8}, \eqref{eq:SF9}, while with respect to the scaling \eqref{eq:SF8} alone neither $\tilde{g}_{\mu\nu}$ nor $\tilde{\chi}$ are invariant.

The fact that some of the scale invariant fields are not dimensionless has an important conceptual consequence. The scale invariance of the effective action does not correspond to a simple change of units for length or mass. Expressing the effective action in terms of dimensionless quantities, it is rather trivial that it remains invariant under a change of length or mass units. This is not the topic here.

\section*{Scale invariant Yang-Mills theory}

The independence of $\Gamma[\tphi]$ of $k$ does not imply that there are no running dimensionless couplings. This running or flow is, however, of a particular type. Any dependence on $k$ is accompanied by a dependence on fields. For example, a running gauge coupling in a scale invariant Yang-Mills theory can occur in the presence of a scalar singlet field $\chi$. The effective running gauge coupling $g(k)$ obeys in one loop order
\begin{equation}
\frac{1}{g^2(k)} = \frac{1}{\bar{g}^2} - \frac{11 N}{48\pi^2} \ln \trho,\quad \tilde{\rho} = \frac{\chi^2}{2k^2},
\label{eq:FE1}
\end{equation}
where we have taken an $\mathrm{SU}(N)$-Yang-Mills theory. 
At fixed $\chi$ the running with $k$ is given by the standard one loop formula
\begin{equation}
k \p_k \frac{1}{g^2} = \frac{11 N}{24\pi^2}.
\label{eq:FE2}
\end{equation}
In the limit $k\to 0$ the scale in the running is effectively replaced by momentum, $k^2\to q^2$. In this limit quantum scale symmetry becomes exact\,\cite{CWQ,CWQS}.
For this version of scale symmetric QCD the UV-cutoff $\Lambda_\mathrm{UV}$ is replaced by a scalar field $\chi$, such that also the confinement scale $\Lambda_\mathrm{QCD}$ is proportional to $\chi$\,\cite{CWQ,SH}.

A lowest order approximation to the effective action of a Yang-Mills theory is given by
\begin{equation}
\Gamma = \frac{1}{4} \int_x \sqrt{g}\sum_z F^z_{\mu\nu} \tilde{Z}_F F^z_{\rho\sigma} g^{\mu\rho} g^{\nu\sigma},
\label{eq:YM1}
\end{equation}
with $F^z_{\mu\nu}$ the field strength for the gauge bosons labeled by $z$. The function $\tilde{Z}_F$ involves the covariant Laplacian
\begin{equation}
\Lapl = -D_\mu D_\nu g^{\mu\nu} = -D_\mu D_\nu k^2 \tilde{g}^{\mu\nu}.
\label{eq:YM2}
\end{equation}
If one is interested in a particular momentum range one may choose $k$ in this range and write
\begin{equation}
\tilde{Z}_F = Z_F z\left( \frac{\Lapl}{k^2} \right).
\label{eq:YM3}
\end{equation}
Both $\sqrt{g}g^{\mu\nu}g^{\rho\sigma} = \sqrt{\tilde{g}} \tilde{g}^{\mu\nu} \tilde{g}^{\rho\sigma}$ and $\Lapl/k^2 = -D_\mu D_\nu \tilde{g}^{\mu\nu}$ involve only scale invariant fields. All possible violations of scale invariance arise therefore from the ``wave function renormalization'' $Z_F$. It is related to the gauge coupling by $Z_F = g^{-2}$.

For standard pure QCD the wave function involves an ultraviolet cutoff as the inverse lattice distance, $Z_F = Z_F(ka)$. The scale $k$ appears explicitly and the theory is not scale invariant. One may introduce renormalized gauge fields $A_{\mathrm{R}\mu}$,
\begin{equation}
A_{\mathrm{R}\mu}(x) = Z_F^\frac{1}{2} (ka) A_\mu(x).
\label{eq:YM4}
\end{equation}
This absorbs the factor $Z_F$ in eq.\,\eqref{eq:YM1}, yielding for constant $z$ a canonical kinetic term. The dependence on $ka$ is then shuffled to the covariant derivative
\begin{equation}
D_\mu = \partial_\mu -iA_\mu^z T_z = \partial_\mu - iZ_F^{-\frac{1}{2}} A_{\mathrm{R}\mu}^z T_z = \partial_\mu - igA_{\mathrm{R}\mu}^z T_z.
\label{eq:YM5}
\end{equation}
There exists no possible choice of gauge invariant fields $\tilde{A}_\mu^z$ for which the dependence on $k$ can be eliminated. Standard QCD is not scale invariant.

For scale invariant Yang-Mills theories the wave function renormalization depends on a scalar field $\chi$ instead of the inverse lattice distance, $Z_F = Z_F(\chi/k)$. All dependence on $k$ is eliminated if one also uses the scale invariant field $\tilde{\chi}$. The gauge field $A_\mu$ is scale invariant.
Since gauge fields carry dimension of mass, this is another example that the scale invariant fields $\tilde{A}_\mu = A_\mu$ are not dimensionless. 

In QCD the running gauge couplings induce a confinement scale $\Lambda_\mathrm{QCD}$. For scale invariant QCD this scale is proportional to $\chi$, $\Lambda_\mathrm{QCD} = \chi/c$. For momenta or $k$ much smaller than the confinement scale the flow of couplings eventually stops and $Z_F$ becomes a constant, $Z_F(\chi/k) \to Z_F(\chi/\Lambda_\mathrm{QCD}) = Z_F(c)$. In this limit scale invariant Yang-Mills theories realize quantum scale symmetry. Of course, the approximation \eqref{eq:YM1} remains no longer valid for momenta near $\Lambda_\mathrm{QCD}$. The fact that the scale $k$ drops out generalizes, however. All mass scales are proportional to $\chi$. A similar setting extends to the electroweak sector of the standard model. The role of $\Lambda_\mathrm{QCD}$ is now assumed by the Fermi scale and the associated masses of the electroweak gauge bosons and fermions. See ref.\,\cite{CWQS} for a more detailed discussion.

\section*{Pregeometry as a gauge theory}

One may formulate\,\cite{CWGG} ``pregeometry'' as a Yang-Mills theory with local gauge symmetry SO(1,\,3) or, in a euclidean version, SO(4). The metric arises in this setting as a composite field, and general relativity corresponds to the effective low energy theory. The six gauge fields $A_\mu^z = A_{\mu mn} = -A_{\mu nm}$ are labeled by a double index $z=(m,n)$, $m,n=0..3$. In addition to fermions $\psi$ we consider a vector field $\tensor{e}{_\mu^m}$, belonging to the four-component vector representation of SO(1,\,3) or SO(4). This vector field will play the role of the vierbein. 

The kinetic term for a Dirac fermion $\psi$ reads
\begin{equation}
\Gamma_{\mathrm{kin}, \psi} = \frac{i}{2} \int e Z_\psi \bar{\psi} \gamma^m D_\mu \psi e_m^\mu + \hc, \quad \bar{\psi} = \psi^\dagger \gamma^0,
\label{eq:SF11}
\end{equation}
with the inverse vierbein $e_m^\mu$ and $e=\det(e_\mu^m)$ replacing $\sqrt{g}$,
\begin{equation}
e_m^\mu e_\mu^n = \delta_m^n,\quad e_m^\mu e_\nu^m = \delta_\nu^\mu,\quad e = \det(e_\mu^m).
\label{eq:SF12}
\end{equation}
The covariant derivative involves the gauge fields
\begin{equation}
D_\mu = \p_\mu -\frac{1}{2} A_{\mu mn} \Sigma^{mn},\quad \Sigma^{mn} = -\frac{1}{4} \comm{\gamma^m, \gamma^n}.
\label{eq:SF13}
\end{equation}
The Dirac matrices obey the usual anticommutation relations
\begin{align}
\begin{split}
\anticomm{\gamma^m, \gamma^n} &= 2\eta^{mn},\quad \\ 
\eta^{mn} = \eta_{mn} &= \diag(-1,1,1,1),
\end{split}
\label{eq:SF14}
\end{align}
and Lorentz indices $m$ are raised and lowered with $\eta^{mn}$ or $\eta_{mn}$.
The euclidean version replaces $\eta_{mn}\to\delta_{mn}$.

The gauge fields are scale invariant, while the scale invariant vierbein and fermion field are given by
\begin{equation}
\tilde{e}_\mu^m = k e_\mu^m,\quad \tilde{\psi} = k^{-3/2} \psi.
\label{eq:SF15}
\end{equation}
The kinetic term \eqref{eq:SF11} is indeed independent of $k$, 
\begin{equation}
\Gamma_{\mathrm{kin}, \psi} = \frac{i}{2} \int \tilde{e} Z_\psi \bar{\tilde{\psi}} \gamma^\mu D_\mu \tilde{\psi} \tilde{e}_m^\mu + \hc ,
\label{eq:SF16}
\end{equation}
provided that $Z_\psi$ is a function of scale invariant fields.
This extends in a straightforward way to Weyl fermions. The scale invariant fermion field $\tilde{\psi}$ is dimensionless, while the scale invariant vierbein carries dimension of mass.

The kinetic term for the gauge bosons,
\begin{equation}
\Gamma_F = \frac{1}{8} \int_x e Z_F F_{\mu\nu,mn} F_{\rho\sigma,pq} g^{\mu\rho} g^{\nu\sigma} \eta^{mp} \eta^{nq},
\label{eq:SF17}
\end{equation}
involves the scale invariant field strength
\begin{equation}
F_{\mu\nu, mn} = \p_\mu A_{\nu mn} - \p_\nu A_{\mu mn} + \tensor{A}{_{\mu m}^p} A_{\nu pn} - \tensor{A}{_{\nu m}^p} A_{\mu pn}.
\label{eq:SF18}
\end{equation}
The metric is a bilinear in the vierbein 
\begin{equation}
g_{\mu\nu} = e_\mu^m e_\nu^n \eta_{mn},\quad g^{\mu\nu} = e_m^\mu e_n^\nu \eta^{mn}.
\label{eq:SF19}
\end{equation}
Similar to the case of other gauge fields \eqref{eq:YM1} the kinetic term \eqref{eq:SF17} does not depend on $k$ once expressed in terms of scale invariant fields. This holds provided that the dimensionless function $Z_F$ only depends on $\trho$ or similar scale invariant quantities.
Both $\Gamma_{\mathrm{kin},\psi}$ and $\Gamma_F$ are invariant under general coordinate transformations (diffeomorphism symmetry).

Gauge symmetry and diffeomorphism symmetry
also allows for a term linear in $F_{\mu\nu,mn}$,
\begin{align}
\begin{split}
\Gamma_\mathrm{R} &= -\frac{1}{8} \int_x F(\chi) F_{\mu\nu,mn} e_\rho^p e_\sigma^q \tensor{\epsilon}{^{\mu\nu\rho\sigma}} \tensor{\epsilon}{^{mn}_{pq}} \\
&= -\frac{1}{2} \int_x e F(\chi) e_m^\mu e_n^\nu \tensor{F}{_{\mu\nu,}^{mn}}\;.
\end{split}
\label{eq:SF20}
\end{align}
Expressed in terms of scale invariant fields and using the dimensionless function $w$ in eq.\,\eqref{eq:SF5} this term becomes
\begin{equation}
\Gamma_\mathrm{R} = -\frac{1}{4} \int_x w F_{\mu\nu,mn} \tilde{e}_\rho^p \tilde{e}_\sigma^q \tensor{\epsilon}{^{\mu\nu\rho\sigma}} \tensor{\epsilon}{^{mn}_{pq}}.
\label{eq:SF21}
\end{equation}
It is independent of $k$ if $w$ is a function involving only scale invariant combinations as $\trho$.

Finally, a gauge invariant kinetic term for the vierbein is constructed from its covariant derivative , the tensor $\tensor{U}{_{\mu\nu}^m}$,
\begin{equation}
\tensor{U}{_{\mu\nu}^m} = D_\mu e_\nu^m = \p_\mu e_\nu^m - \tensor{\Gamma}{_{\mu\nu}^\lambda}(e) e_\lambda^m + \tensor{A}{_\mu^m_n} e_\nu^n,
\label{eq:SF22}
\end{equation}
where
the Levi-Civita connection
$\tensor{\Gamma}{_{\mu\nu}^\lambda}$ depends on the vierbein via eq.\,\eqref{eq:SF19}
\begin{equation}
\tensor{\Gamma}{_{\mu\nu}^\lambda}(e) = \frac{1}{2}g^{\lambda\rho} \left( \p_\mu g_{\nu\rho} + \p_\nu g_{\mu\rho} - \p_\rho g_{\mu\nu} \right).
\label{eq:SF22A}
\end{equation}
The kinetic term involves functions $m^2(\chi)$ and $m_\mathrm{V}^2(\chi)$ with dimension mass squared,
\begin{equation}
\Gamma_U = \frac{1}{4} \int_x e \tensor{U}{_{\mu\nu}^m} \tensor{U}{_{\rho\sigma}^n} \eta_{mn} \left( m^2 g^{\mu\rho} g^{\nu\sigma} + m_\mathrm{V}^2 g^{\mu\nu} g^{\rho\sigma} \right).
\label{eq:SF23}
\end{equation}
In terms of the scale invariant vierbein it reads
\begin{equation}
\Gamma_U = \frac{1}{4} \int_x \tilde{e} \tensor{\tilde{U}}{_{\mu\nu}^m} \tensor{\tilde{U}}{_{\rho\sigma}^n} \eta_{mn} \left( \tilde{m}^2 \tilde{g}^{\mu\rho} \tilde{g}^{\nu\sigma} + \tilde{m}_\mathrm{V}^2 \tilde{g}^{\mu\nu} \tilde{g}^{\rho\sigma} \right)
\label{eq:SF24}
\end{equation}
with dimensionless functions
\begin{equation}
\tilde{m}^2 = \frac{m^2}{k^2},\quad \tilde{m}_\mathrm{V}^2 = \frac{m_\mathrm{V}^2}{k^2}.
\label{eq:SF25}
\end{equation}
Independence of $k$ follows if $\tilde{m}^2$ and $\tilde{m}_\mathrm{V}^2$ are functions of scale invariant fields.

We assume that all dimensionless functions as $u$, $w$, $K$, $Z_F$, $Z_\psi$, $\tilde{m}^2$ or $\tilde{m}_\mathrm{V}^2$, only depend on dimensionless fields or invariants as $\trho$. As a consequence, no intrinsic length or mass scale is present in the effective action. 
The model of pregeometry based on $\Gamma_U + \Gamma_F + \Gamma_R$ exhibits fundamental scale invariance.
In terms of the canonical fields $g_{\mu\nu}$, $\chi$, $e_\mu^m$, $A_\mu$, $\psi$ the only mass scale appearing in the effective action is the ``renormalization scale'' $k$. It is only introduced by the transition from the scale invariant fields to canonical fields.

From the Levi-Civita-connection $\tensor{\Gamma}{_{\mu\nu}^\rho}(e)$ 
a curvature tensor $R_{\mu\nu\rho\sigma}(e)$ is defined as a function of the vierbein in the standard way. 
The field strength
$F_{\mu\nu mn}$ and $R_{\mu\nu\rho\sigma}(e)$ are related by the commutator of covariant derivatives of the vierbein\,\cite{CWGG}
\begin{align}
\begin{split}
\comm{D_\mu,D_\nu} e_\rho^m &= \tensor{F}{_{\mu\nu}^m_n} \tensor{e}{_\rho^n} - \tensor{R}{_{\mu\nu}^\sigma_\rho}(e) \tensor{e}{_\sigma^m} \\
&= D_\mu \tensor{U}{_{\nu\rho}^m} - D_\nu \tensor{U}{_{\mu\rho}^m} 
= \tensor{V}{_{\mu\nu\rho}^m}.
\end{split}
\label{eq:SF28}
\end{align}
If the tensor $\tensor{V}{_{\mu\nu\rho}^m}$ vanishes the field strength can be identified with the curvature tensor,
\begin{equation}
e_\rho^m e_\sigma^n F_{\mu\nu mn} = R_{\mu\nu\rho\sigma}(e).
\label{eq:SF27}
\end{equation} 
In this case the term \eqref{eq:SF20} equals the term proportional to the curvature scalar $R(e)$ in eq.\,\eqref{eq:I3}. 

This is the way how standard Riemannian geometry with the Einstein-Hilbert action can be recovered as a low energy limit. Due to the quadratic term \eqref{eq:SF23} the field equations lead for low momenta to the approximate solution $\tensor{U}{_{\mu\nu}^m}=0$, and therefore $\tensor{V}{_{\mu\nu\rho}^m}=0$.
For $\tensor{U}{_{\mu\nu}^m}=0$ the gauge field equals the usual spin connection, $A_{\mu mn} = \omega_{\mu mn}$, as given by
\begin{align}
\begin{split}
\omega_{\mu np} &= -e_\mu^m \left( \Omega_{mnp} - \Omega_{npm} + \Omega_{pmn} \right), \\
\Omega_{mnp} &= -\frac{1}{2} \left( e_m^\mu e_n^\nu - e_n^\mu e_m^\nu \right) \p_\mu e_{\nu p}. 
\end{split}
\label{eq:SF26}
\end{align}

For vanishing $\tensor{V}{_{\mu\nu\rho}^m}$ the term \eqref{eq:SF17} involves the squared Riemann tensor. This invariant contains four derivatives of the vierbein. The higher order derivatives only appear through the identification \eqref{eq:SF27}. Similar to the curvature tensor, there are other possible contractions of the field strength. We define
\begin{equation}
F_{\mu m} = F_{\mu\nu mn} e^{n \nu},\quad F = F_{\mu m} e^{m \mu},
\label{eq:SF29}
\end{equation}
and generalize the term \eqref{eq:SF17} to
\begin{align}
\begin{split}
\Gamma_A = \int_x e &\left\{ \frac{Z_F}{8} F_{\mu\nu mn} F^{\mu\nu mn} \right. + \left.\frac{A}{2} F_{\mu m} F^{\mu m} + \frac{B}{2} F^2 \right\}.
\end{split}
\label{eq:SF30}
\end{align}
For $\tensor{V}{_{\mu\nu\rho}^m}=0$ this generates corresponding four-derivative invariants formed from the Riemann tensor $R_{\mu\nu\rho\sigma}(e)$.

In contrast to four-derivative gravity the high-momentum limit of this version of pregeometry has standard propagators for all fields, as common for actions involving up to two derivatives. The ghost instability of the graviton propagator in four-derivative gravity is expected to be an artifact of the truncation of a polynomial expansion in the number of derivatives\,\cite{PLAW}. The graviton propagator in flat space multiplies a momentum dependent function $G_\mathrm{grav}(q^2)$ with an appropriate projector on the traceless transverse tensor mode. For constant $Z_F=Z$, $A=B=0$, and constant $m^2$, $M^2$ the inverse graviton propagator reads
\begin{align}
\label{eq:49A}
&G_\mathrm{grav}^{-1}(q^2) = \frac{m^2}{8} \left\{\vphantom{\sqrt{\frac{0}{0}}} (Z+1)q^2 + m^2 - M^2 \right. \\
\nonumber
&\quad - \left.\sqrt{[(Z-1)q^2 +m^2 -M^2]^2 + 4 \frac{q^2}{m^2}(m^2-M^2)^2 } \right\}. 
\end{align}
It obtains by diagonalization of the inverse propagator matrix in the transverse traceless sector. For 
\begin{equation}
m^2>0,\quad M^2>0,\quad 0<Z<\frac{M^2}{m^2} \left( 1-\frac{M^2}{m^2} \right)
\label{eq:49B}
\end{equation}
this propagator has a single pole in the complex $q^2$-plane at $q^2=0$. Analytic continuation from euclidean signature ($q^2 \geq 0$) to Minkowski signature ($q^2 = -q_0^2 + \vec{q}^2$) can be performed. In the complex $q_0$-plane for Minkowski signature branch cuts occur on the real axis for $q_0^2 > |q_\mathrm{c}|^2 + \vec{q}^2$, $|q_\mathrm{c}|^2>0$. This type of model can be considered as a valid candidate for an ultraviolet completion of quantum gravity.

Running couplings $Z_F(\tilde{\rho})$, $\tilde{m}^2(\tilde{\rho})$ are compatible with fundamental scale invariance.
The scaling solutions for $Z_F(\tilde{\rho})$, $\tilde{m}^2(\tilde{\rho})$ could become simple in the ultraviolet limit $\tilde{\rho}\to 0$, or $k\to\infty$ at fixed $\chi$. If $Z_F(\tilde{\rho})$ diverges in this limit the gauge sector is asymptotically free. A constant value $\tilde{m}^2(\tilde{\rho}\to 0) = \tilde{m}_0^2$ can be absorbed by a rescaling of fields and is no free parameter. An interesting limit arises if $M^2/m^2$ reaches zero for $\tilde{\rho}\to 0$. We observe that gauge fields and vierbein may arise as composites in an even more fundamental pregeometry as spinor gravity\,\cite{CWLSG}. Regulated on a lattice, such a theory can indeed be formulated without any scale, giving a strong motivation for fundamental scale invariance.

\section*{Flow equation}

A convenient method for the investigation of running couplings in theories with fundamental scale invariance is functional renormalization for the effective average action\,\cite{CWFE}.
The flow equation is most conveniently formulated in terms of canonical fields $\varphi$ and we will discuss the version with scale invariant fields $\tilde{\varphi}$ subsequently.
We define the effective average action or flowing action $\Gamma_k[\phi]$ similarly to eq.\,\eqref{eq:NS1} by adding an infrared cutoff term $\Delta_k[\chi]$\,\cite{CWGIF}, 
\begin{equation}
\Gamma_k[\phi] = -\ln \left( Z_k[\phi] \right) - C_k[\phi],
\label{eq:FE4}
\end{equation}
where the $k$-dependent partition function reads
\begin{align}
\begin{split}
Z_k[\phi] = \int \mathcal{D}\chi \exp &\left\{
-S[\phi+\chi] -\Delta_k[\chi;\phi] \phantom{\frac{0}{0}} \right. \\
&+ \left. \int_x \left( \frac{\p\Gamma_k}{\p\phi} + L_k[\phi] \right) \chi \right\}.
\end{split}
\label{eq:FE5}
\end{align}
The cutoff term is bilinear in the fluctuation fields $\chi$ and may depend on the macroscopic fields $\phi$,
\begin{equation}
\Delta_k[\chi; \phi] = \frac{1}{2} \int_x \chi_i(x) R_{k,ij} (-D^2;\phi) \chi_j(x).
\label{eq:FE6}
\end{equation}
The covariant Laplacian $D^2$ (or some similar operator) is formed with the macroscopic fields $\phi$, such that the cutoff \eqref{eq:FE6} can be made invariant under local gauge transformations.
The functionals $C_k[\phi]$ and $L_{k,i}(x)[\phi]$ can be used for optimization and will be discussed later.
The fields $\varphi$ and $\chi$ stand here for arbitrary bosonic fields, including gauge fields, vierbein or the metric, with a well known generalization for fermions.

The dimension of the cutoff function $R_k$ is dictated by the dimension of the fields $\chi_i(x)$. For the example of scalars the dimension of $R_k$ is mass squared, and we introduce a dimensionless function $r_k$
\begin{equation}
R_k = ek^2 r_k \left( -\frac{D^2}{k^2}, \frac{\phi}{k} \right).
\label{eq:FE7}
\end{equation}
This can be generalized to fields with other dimensions. We require that $R_k$ vanishes for $k\to 0$, such that we recover the effective action \eqref{eq:NS1} in this limit. For $k\neq 0$ the quadratic term \eqref{eq:FE6} acts as an infrared cutoff. For high momenta, corresponding to large values of $-D^2/k^2$, the cutoff function is chosen to vanish rapidly, such that the functional integral over fluctuations with momenta much larger than $k$ is not affected. For $k\to\infty$ the quadratic term $\sim\Delta_k$ dominates the functional integral which becomes effectively Gaussian. In this limit one typically has $\lim_{k\to\infty} \Gamma_k[\phi] \approx S[\phi]$. The effective average action interpolates between the classical action for $k\to\infty$ and the quantum effective action for $k\to 0$. The fluctuation effects that map $S[\phi]$ to $\Gamma[\phi]$ are taken into account in continuous steps.

We consider cutoff functions that depend on the macroscopic fields\,\cite{CWGIF}. This permits us to maintain local gauge symmetries by employing covariant derivatives constructed from the connection which involves the macroscopic vierbein (or metric) or macroscopic gauge fields. In particular, diffeomorphism symmetry requires that $R_k$ is proportional to $e=\sqrt{g}$. The price to pay for maintaining gauge symmetry are corrections to the flow equation that may be minimized by suitable optimization functionals $C_k[\phi]$, $L_k[\phi]$. These optimization functionals vanish for $R_k=0$ and therefore for $k\to 0$, while for $k\to\infty$ one has vanishing $C_k$ and finite $L_k$.
The use of the macroscopic fields (instead of the often used independent ``background fields'') in the cutoff induces some particular features that we discuss briefly. For cutoffs not involving the macroscopic fields the correction terms $C_k[\varphi]$, $L_k[\varphi]$ are absent and one recovers the standard formulation of the effective average action and associated flow equation. We will define these correction terms in the following. They will not be needed for practical calculations.

Expectation values are computed in the presence of $\Delta_k$, 
\begin{align}
\begin{split}
\braket{A} = \inv{Z_k} \int \mathcal{D}\chi A[\chi] \exp &\left\{ -S[\phi+\chi] - \Delta_k[\chi] \phantom{\frac{\p\Gamma_k}{\p\phi}} \right. \\
&\left. \quad+ \int_x \left( \frac{\p\Gamma_k}{\p\phi} + L_k \right) \chi \right\}.
\end{split}
\label{eq:FE8}
\end{align}
They therefore depend on $k$. We may consider the family of effective average actions $\Gamma_k[\phi]$ for different $k$ as a family of different models, labeled by $k$. The models apparently differ by their infrared cutoffs and have the same high momentum behavior.
We can shift the integration to $\sigma=\phi+\chi$,
\begin{align}
\begin{split}
\braket{A} = \inv{Z_k} \int &\mathcal{D}\sigma A[\sigma] \exp \left\{ -S[\sigma] - \Delta_k[\sigma-\phi; \phi] \phantom{\frac{-}{-}} \right. \\
&+ \left.\int_x \left( \frac{\p\Gamma_k}{\p\phi}+L_k \right) (\sigma-\phi) \right\},
\end{split}
\label{eq:FE9}
\end{align}
where
\begin{align}
\begin{split}
Z_k = \int &\mathcal{D}\sigma \exp \left\{ -S[\sigma] - \Delta_k[\sigma-\phi; \phi] \phantom{\frac{-}{-}} \right. \\
&+ \left.\int_x \left( \frac{\p\Gamma_k}{\p\phi}+L_k \right) (\sigma-\phi) \right\}.
\end{split}
\label{eq:FE10}
\end{align}

We first want to 
relate the macroscopic field $\phi_i(x)$ to the expectation value of the microscopic field $\bar{\sigma}_i(x) = \braket{\sigma_i(x)}$.
For this purpose we take
the functional derivative of $\Gamma_k[\phi]$,
\begin{align}
\begin{split}
&\frac{\p\Gamma_k}{\p\phi_i(x)} = \braket{\frac{\p}{\p\phi_i(x)} \Delta_k[\sigma-\phi; \phi] } \\
&- \int_y \left( \frac{\p^2\Gamma_k}{\p\phi_i(x) \p\phi_i(y)} + \frac{\p L_{k,j}(y)}{\p\phi_i(x)} \right) (\bar{\sigma}_j(y) - \phi_j(y)) \\
&+ \frac{\p\Gamma_k}{\p\phi_i(x)} + L_{k,i}(x) - \frac{\p C_k}{\p\phi_i(x)}.
\end{split}
\label{eq:FE11}
\end{align}
Here the $\phi$-derivative of $\Delta_k$ in the expectation value (first term on \rhs) is performed under the integral at fixed $\sigma$. 
One finds
\begin{align}
\begin{split}
\int_y \left( \frac{\p^2\Gamma_k}{\p\phi_i(x) \p\phi_j(y)} + \frac{\p L_{k,j}(y)}{\p\phi_i(x)} \right) (\bar{\sigma}_j(y)-\phi_j(y)) = -K_i(x),
\end{split}
\label{eq:FE12}
\end{align}
where
\begin{align}
\begin{split}
K_i(x) = \braket{\frac{\p}{\p\phi_i(x)} \Delta_k[\sigma-\phi; \phi] } + L_{k,i}(x) - \frac{\p C_k}{\p\phi_i(x)}
\end{split}
\label{eq:FE13}
\end{align}
vanishes for $k=0$, $\Delta_k=0$. For $k=0$, $L_0=0$, $C_0=0$, the macroscopic field equals the expectation value of the microscopic field $\phi=\bar{\sigma}$, as for the usual construction of the effective action by a Legendre transform of the Schwinger functional. 

For $k\neq 0$ a non-zero $K_i(x)$ can arise from a possible dependence of the cutoff function $eR_{k,jl}$ on $\phi$,
\begin{align}
\begin{split}
K_i(x) = &\frac{1}{2} \tr \left\{ \frac{\p R_k}{\p\phi_i(x)} G \right\} - (R_k)_{ij} \bar{\chi}_j(x) \\
&+ L_{k,i}(x) - \frac{\p C_k}{\p\phi_i(x)}.
\end{split}
\label{eq:FE14}
\end{align}
Here $G$ is the matrix of two-point functions
\begin{equation}
G_{jl}(x,y) = \braket{\chi_j(x) \chi_l(y)}.
\label{eq:FE15}
\end{equation}
For $\bar{\chi}=0$ it is the connected two-point function or the propagator for $\sigma$.
We will choose $L_k$ such that $K=0$ and $\bar{\chi}=\bar{\sigma}-\varphi=0$. Then the macroscopic field $\varphi$ equals the expectation value of the microscopic field $\bar{\sigma}$.
In the trace in eq.\,\eqref{eq:FE14} we consider the IR-cutoff function as a matrix
\begin{equation}
R_{k,jl}(x,y) = \delta(x-y) R_{k,jl} \left( -D_y^2; \phi(y) \right),
\label{eq:FE16}
\end{equation}
and we have assumed for simplicity that $R_k$ is symmetric. For $k\neq 0$ the first term on the \rhs\ of eq.\,\eqref{eq:FE14} does not vanish
if the cutoff depends on the macroscopic field. 
The optimization terms $L_k$ and $C_k$ are used to cancel this term. For any choice of $C_k$ this defines the functional $L_{k,i}(x)$
by setting $\bar{\chi}=0$, $K=0$ in eq.\,\eqref{eq:FE14},
\begin{equation}
L_{k,i}(x) = -\frac{1}{2} \tr \left\{ \frac{\partial R_k}{\partial \varphi_i(x)} G \right\} + \frac{\partial C_k}{\partial \varphi_i(x)}.
\end{equation} 
With this choice one has $\braket{\sigma_i(x)} = \phi_i(x)$, $\bar{\chi}_i(x)=0$ for all $k$, despite the dependence of $R_k$ on the macroscopic fields.
Also $G$ is the connected two-point function of the microscopic fields

We will next determine the correction term $C_k$ such that the flow equation for $\Gamma_k$ takes the usual simple form. This is achieved by relating $G$ to the second functional derivative $\Gamma_k^{(2)}$. The exact flow equation for the effective average action obtains by taking a $k$-derivative of eqs.\,\eqref{eq:FE4}, \eqref{eq:FE5}
\begin{equation}
\p_k\Gamma_k[\phi] = \frac{1}{2} \tr \left\{ (\p_k R_k)G\right\} - \p_k C_k[\phi].
\label{eq:FE17}
\end{equation}
This simple form employs $\bar{\chi}=0$. We write
\begin{equation}
G = \inv{(\Gamma_k^{(2)} + R_k)} + \Delta_k G,
\label{eq:FE18}
\end{equation}
with $\Gamma_k^{(2)}$ the matrix of second functional derivatives of $\Gamma_k$. The correction term $\Delta_k G$ can be computed as in ref.\,\cite{CWGIF}. It vanishes if $R_k$ is independent of the macroscopic fields $\phi$. We arrive at the flow equation
\begin{equation}
\p_k\Gamma_k[\phi] = \frac{1}{2} \tr \left\{ (\p_k R_k) \inv{(\Gamma_k^{(2)} +R_k)} \right\} + B_k[\phi].
\label{eq:FE19}
\end{equation}
The correction term,
\begin{equation}
B_k[\phi] = \frac{1}{2} \tr \left\{ (\p_k R_k) \Delta_k G \right\} -\p_k C_k,
\label{eq:FE20}
\end{equation}
vanishes for a suitable choice of $C_k$. For $R_k$ independent of the macroscopic fields one has $C_k=0$.
The condition $B_k[\varphi]=0$ defines a functional differential equation for $C_k[\varphi]$, with initial condition $C_0[\varphi]=0$. We only need the existence of the solution, for which we see no obstruction. We assume the existence of a solution and define $C_k[\varphi]$ accordingly.

In summary, by a suitable choice of $L_k$ and $C_k$ in the definition of $\Gamma_k$ we arrive at an effective average action that obeys the standard exact flow equation\,\cite{CWFE}. Furthermore, the macroscopic field $\phi$ equals the expectation value of the microscopic field $\braket{\sigma}$ for all $k$
and the $k$-dependent propagator matrix $G$ equals the connected two-point function for $\sigma$. These properties, together with $\Gamma_{k\to\infty}[\varphi]\approx S[\varphi]$, are sufficient for all practical purposes. The flow equation, together with the initial condition for $k\to\infty$, can be used for an alternative definition of the theory, without invoking the functional integral explicitly. The latter describes then the formal solution of the functional differential flow equation. 
The exact form of the flow equation will actually not be crucial for our discussion
of fundamental scale invariance. 
What is important is the existence of a flow equation that can account for running couplings.

\section*{Flow equation for scale invariant fields}

In the preceding discussion we have defined the effective average action \eqref{eq:FE4}, \eqref{eq:FE5} as a functional of the canonical fields. The flow equation describes the variation with the scale $k$ for fixed canonical fields $\phi$. Let us now express $\Gamma_k$ as a functional of the scale invariant fields. For the classical action $S$ this has been discussed previously. With $\phi$ and $\chi$ scaling in the same way as $\sigma$ the action $S[\tphi+\tchi]$ 
of a theory with fundamental scale invariance
does not depend on the scale $k$. Since the relation between $\tchi_i$ and $\chi_i$ is only a $k$-dependent factor, the functional measures $\int \mathcal{D}\chi$ and $\int \mathcal{D}\tchi$ differ only by a $k$-dependent but field independent factor. This only results in an irrelevant additive constant for $\Gamma_k$.

For the infrared cutoff $\Delta_k$ we choose the same $k$ for the transition to scale invariant fields as the one that appears in $R_k$. As a result the cutoff term becomes independent of $k$ once it is expressed in terms of scale invariant fields. This may be demonstrated by the cutoff \eqref{eq:FE7} for scalar fields. With $e=k^{-4}\tilde{e}$ one has
\begin{equation}
R_k = k^{-2} \tilde{e} r_k(-\tilde{D}^2;\tphi),
\label{eq:EA1}
\end{equation}
such that $r_k$ no longer involves $k$.
Here we use
\begin{align}
\begin{split}
-D^2 = -g^{\mu\nu} D_\mu D_\nu &= -k^2 \tilde{g}^{\mu\nu} D_\mu D_\nu = -k^2 \tilde{D}^2, \\
-\frac{D^2}{k^2} &= -\tilde{D}^2,\quad \frac{\phi}{k} = \tphi.
\end{split}
\label{eq:EA2}
\end{align}
The factor $k^{-2}$ in eq.\,\eqref{eq:EA1} is canceled by $\chi^2 = k^2 \tchi^2$, resulting in $\Delta_k$ becoming independent of $k$
\begin{equation}
\Delta_k = \frac{1}{2} \int_x \tilde{e} \tilde{\chi}^\mathrm{T} r(-\tilde{D}^2; \tilde{\varphi}) \tilde{\chi}.
\label{eq:70A}
\end{equation}
This holds similarly for fields with other scaling dimensions if the prefactor multiplying $r_k$ involves besides $e$ only powers of $k$, multiplied by possible functions of scale invariant fields.

Finally, $\p\Gamma_k/\p\chi_i(x)$ scales inversely to $\chi_i(x)$ and similar for $L_{k,i}(x)$. Up to an irrelevant multiplicative factor one finds
\begin{align}
\begin{split}
Z_k[\tilde{\varphi}] = \int \mathcal{D}\tchi \exp &\left\{ -S[\tphi+\tchi] -\Delta_k[\tchi; \tphi] \phantom{\frac{I}{I}} \right. \\
&\left. + \int_x \left( \frac{\p\Gamma_k}{\p\tphi} + \tilde{L}_k \right) \tchi \right\}.
\end{split}
\label{eq:EA3}
\end{align}
Since $\Delta_k[\tchi; \tphi]$ no longer involves the scale $k$, one finds $Z_k[\tphi]$ and $\Gamma_k[\tphi]$ independent of $k$. This requires the optimization functionals $\tilde{L}_k[\tphi]$ and $C_k[\tphi]$ to be independent of $k$ once expressed in terms of scale invariant fields. This is self-consistent.

We arrive at an important conclusion: The effective average action does no longer involve the scale $k$ if it is written as a functional of the scale invariant fields. The whole family of apparently different effective average actions for different $k$ describes actually the same model. The difference between the different members of the family is only due to the use of different canonical fields, all corresponding to the same scale invariant fields, but using different $k$ for the scaling. As an immediate consequence, the flow with $k$, evaluated for fixed scale invariant fields, vanishes,
$\p_k \Gamma_k[\tphi] = 0$.

This is precisely the setting \eqref{eq:I8} for a theory with fundamental scale invariance. The introduction of the infrared cutoff has not changed this.

How can the effective average action for scale invariant fields describe a running of couplings despite the fact that no scale $k$ appears anymore? The functional integral still contains an infrared cutoff term. It is now a fixed term, corresponding to setting $k=1$. The flow occurs now in field space. Changing the value of $\tphi$ indeed amounts for fixed $\phi$ to a change in $k$. The average effective action is a fixed functional, and the flow equation describes what happens if we rescale the field values according to the appropriate dimension. For scalar fields, flowing towards the infrared corresponds to an increase of $\trho$.

\section*{Quantum scale symmetry}

Dilatation transformations or global scale transformations are rescalings of the canonical fields $\phi$ at \textit{fixed} $k$. The possible scale symmetry associated to these transformations can be violated
for theories with fundamental scale invariance.
The effective action $\Gamma_k[\phi]$ is not independent of $k$, and not invariant under rescalings of canonical fields $\phi$ at fixed $k$. It is only invariant under simultaneous rescalings of $\phi$ and $k$.
Quantum scale symmetry is associated to the invariance of the effective action under dilatations or global scale transformations at fixed $k$. If this symmetry is spontaneously broken 
by a non-zero scalar field $\chi$
one expects a Goldstone boson. For a theory with fundamental scale invariance the 
presence of the
scale $k$ 
in the effective action
appears as a dilatation anomaly which typically turns the Goldstone boson to a pseudo-Goldstone boson. Fundamental scale invariance of a theory is a property rather than a global symmetry that could be broken spontaneously. Under simultaneous scale transformations of $k$ and the canonical fields $\varphi$ the scaling fields $\tilde{\varphi}$ are simply invariant.

For an illustration we discuss for a theory with fundamental scale invariance an effective scalar potential of the form
\begin{align}
\begin{split}
\sqrt{g}U &= \sqrt{\tilde{g}}u = \frac{1}{8}\sqrt{\tilde{g}} \tilde{\lambda}(\tchi) \left( \tchi^2 - \kappa \right)^2 \\
&= \frac{1}{8}\sqrt{g} \tilde{\lambda} \left(\frac{\chi}{k}\right) \left(\chi^4 -2\kappa k^2\chi^2 + \kappa^2 k^4 \right).
\end{split}
\label{eq:I10}
\end{align}
The minimum occurs for $\chi_0^2 = \kappa k^2$, and the mass term $m^2= \p^2U/\p \chi^2$ does not vanish. There is no Goldstone boson despite the fact that no intrinsic scale is present.
This generalizes to other forms of the scaling potential, as the characteristic non-polynomial potentials found in scaling solutions\,\cite{ESPA}.

We have to distinguish between scale invariance, which means the absence of intrinsic mass scales, and scale or dilatation symmetry, which means invariance under rescaling of canonical fields at fixed $k$. 
The potential \eqref{eq:I10} is scale invariant, but not dilatation symmetric. Scale invariance is realized if $k$ is the only scale appearing in the effective action, and if $k$ can be eliminated by a transition to scaling fields. The criterion for scale invariance is eq.\,\eqref{eq:I8}. Scale invariance would be violated if we introduce in eq.\,\eqref{eq:I10} an additional mass parameter $\mu^2$ by a term $\sqrt{g}\mu^2\chi^2$. From the point of view of flow equations this corresponds to a relevant parameter for a deviation from a scaling solution. Quantum scale symmetry or dilatation symmetry requires that no scale is present at all in the effective action, even not $k$. Our example \eqref{eq:I10} shows explicitly that a scale invariant effective action can violate quantum scale symmetry.

There are particular limits for which quantum scale symmetry becomes exact for scale invariant theories. For these limits the effective action becomes invariant under global scalings of the canonical fields. In particular, they concern the limiting behavior for $\tchi\to 0$ or $\tchi\to\infty$. If for $\tchi\to 0$ the effective action $\Gamma[\tchi]$ reaches a well defined limit, the dimensionless couplings $g(\tchi)$ reach limits $g_*$ that do no longer depend on $\tchi$. At the same time, they do not depend on $\chi$ and on $k$. Their flow with $k$ stops -- the couplings approach a fixed point. 
Quantum scale symmetry is realized at a fixed point if all couplings are dimensionless. Since 
$k$ is the only scale in a theory with fundamental scale invariance, and it drops out at the fixed point,
no parameter with dimension of mass or length is present in $\Gamma[\chi]$ anymore.
At fixed $\chi$ the limit $\tchi\to 0$ corresponds to a diverging ``renormalization scale'' $k\to\infty$. This limit is an ultraviolet (UV) fixed point. 

If in the limit $\tchi\to\infty$ the effective action also reaches a well defined limit, dimensionless couplings become again independent of $\tchi$. This corresponds to an infrared (IR) fixed point, since for fixed $\chi$ the renormalization scale $k$ reaches zero. 
In a scale invariant setting the UV- and IR-fixed points are in the first instance fixed points in the dependence of couplings on the scale invariant field $\tchi$. This can translate to the independence of $k$ and global scale symmetry. For the IR-fixed point one has $\chi\to\infty$ at fixed $k$, such that the exact quantum scale symmetry is spontaneously broken
by the non-zero value of $\chi$. Particles can be massive with masses $\sim\chi$. A massless Goldstone boson is predicted. For the UV-fixed point fixed $k$ corresponds to $\chi\to 0$. A possible global scale symmetry is not spontaneously broken and all particle masses go to zero in this case.

In the presence of both an UV-fixed point for $\tilde{\chi}\to 0$ and an IR-fixed point for $\tilde{\chi}\to\infty$ the intermediate values of $\tilde{\chi}$ describe a crossover between the two fixed points. We recall here that the choice of metric and scalar fields is not unique. The metric frame can be changed by suitable field redefinitions. Quantum scale symmetry is often only seen for an appropriate choice of fields as, for example, the primordial flat frame\,\cite{PFF1,PFF2} for the UV-fixed point.

Cosmology can be described by a crossover\,\cite{CWQIM,CWQS}, where $\tilde{\chi}$ increases from zero in the infinite past to infinity in the infinite future. Inflation is the early period near the UV-fixed point, while the present cosmological epoch is already close to the IR-fixed point with very large $\tchi$. The pseudo-Goldstone boson is the cosmon, which is responsible for dynamical dark energy. According to eq.\,\eqref{eq:I2}, the present dark energy density in units of the Planck mass is tiny, $\lambda\sim\tchi^{-4}$, for large values of the dimensionless scale invariant field $\tchi$.

\section*{Discussion}

We have investigated theories with fundamental scale invariance. 
Scale invariant fields are related to canonical fields by an arbitrary renormalization scale $k$. We choose $k$ to be the effective infrared cutoff in the
formulation of the effective average action. 
An ultraviolet or microscopic scale can be given by some inverse lattice distance $\inv{a}$ or similar. 
For theories with fundamental scale invariance a continuum limit $ka\to 0$ exists
with \textit{fixed} microscopic couplings at momenta $a^{-1}$ or fixed lattice couplings.
In this case the effective average action 
can be written
as a functional of the scale invariant fields which remains well defined in the continuum limit.

Theories with fundamental scale invariance have a close connection to quantum scale symmetry. 
Whenever all dimensionless couplings become independent of the scale invariant fields, and therefore independent of $k$ for fixed canonical fields, exact quantum scale symmetry is realized. In particular, if 
an infrared fixed point is reached for $k\to 0$ at fixed canonical fields, one recovers exact quantum scale symmetry. For a given scaling solution, and a given cosmological solution of the field equations derived from the corresponding effective action, one can infer the value of $k$
which is relevant for the present cosmological epoch. In
standard particle physics units,
fixed by the present value of the Planck mass,
it turns out to be $k\approx 10^{-3}\,\mathrm{eV}$. 
This value is much smaller than the characteristic scales in particle physics. Fundamental scale invariance predicts for the present cosmological epoch a standard model with quantum scale symmetry.
An exception may be neutrino masses.

Furthermore, for
many observations there are physical cutoffs that stop effectively the flow of couplings. We may associate such physical cutoffs with some squared momentum $q^2$. 
For $q^2 \gg k^2$ one can effectively replace $k^2$ by $q^2$ in the scaling solutions
in a very good approximation.
With this replacement the effective average action corresponding to the scaling solution exhibits exact quantum scale symmetry.

Theories with fundamental scale invariance have a very high predictive power, much stronger than arbitrary renormalizable theories. General renormalizable theories, both asymptotically free or asymptotically safe, have free parameters corresponding to the so called relevant parameters for small deviations of the flow from an ultraviolet fixed point. Theories with fundamental scale invariance correspond to exact scaling solutions of the flow equations. All relevant parameters vanish, and are therefore not available as free parameters for an interpretation of observations. If there is a unique scaling solution, theories with fundamental scale symmetry contain no free parameters. Free parameters can only arise if there exist families of scaling solutions, with parameters distinguishing between different members of such families.

The existence of scaling solutions is already highly non-trivial. It guarantees that a theory is ``renormalizable'' or ``ultraviolet complete''. These scaling solutions are all what is needed for theories with fundamental scale invariance. In contrast to general renormalizable theories no deviations from the scaling solution due to relevant parameters need to be studied 
for theories with fundamental scale invariance.

In the presence of quantum gravity the scaling solutions often have properties that are not familiar in perturbation theory for particle physics. For example, the effective scalar potential may reach a constant for large values of the fields\,\cite{ESPA}. Together with an effective Planck mass increasing
proportional to a scalar field $\tilde{\chi}$
this solves the cosmological constant problem asymptotically, without any tuning of parameters. What is usually a tuning of parameters becomes the statement that for scaling solutions the effective potential becomes constant for large $\tchi$ instead of increasing $\sim\tchi^4$. It is well conceivable that other perturbative tuning problems as the gauge hierarchy could find a solution by properties of scaling solutions. It is highly unlikely that families of scaling solutions with twenty or more free parameters exist. As a consequence, many renormalizable couplings of the standard model of particle physics become predictable. 
All predictions from the renormalizability of quantum gravity (asymptotic safety) carry over to theories with fundamental scale invariance. A prime example is the prediction of the mass of the Higgs boson\,\cite{SW} or the mass of the top quark\,\cite{ESPA} for a given observed mass of the Higgs boson.

Fundamental scale invariance is a new theoretical construction principle beyond renormalizability. The existence of a continuum limit at fixed microscopic couplings is a very natural setting for a fundamental theory. The required existence of scaling
solutions could be sufficiently restrictive to 
qualify theories with fundamental scale invariance as
candidates for the quest of a unified fundamental theory of physics.

 
 


\bibliography{refs}

\end{document}